\documentclass[aps,showpacs,twocolumn,draft,eqsecnum]{revtex4}

\usepackage{epsfig}
\usepackage{latexsym}

\def \be{\begin{equation}}
\def \ee{\end{equation}}
\def \bea{\begin{eqnarray}}
\def \eea{\end{eqnarray}}

\begin{document}


\title{A hyperbolic slicing condition adapted to Killing fields and
densitized lapses}


\author{Miguel Alcubierre}
\email{malcubi@nuclecu.unam.mx}
\author{Alejandro Corichi}
\email{corichi@nuclecu.unam.mx}
\author{Jos\'e A. Gonz\'alez}
\email{cervera@nuclecu.unam.mx}
\author{Dar\'{\i}o N\'u\~nez}
\email{nunez@nuclecu.unam.mx}
\author{Marcelo Salgado}
\email{marcelo@nuclecu.unam.mx}

\affiliation{Instituto de Ciencias Nucleares, Universidad Nacional
Aut\'onoma de M\'exico, A.P. 70-543, M\'exico D.F. 04510, M\'exico.}


\date{March, 2003.}


\begin{abstract}
We study the properties of a modified version of the Bona-Masso family
of hyperbolic slicing conditions.  This modified slicing condition has
two very important features: In the first place, it guarantees that if
a spacetime is static or stationary, and one starts the evolution in a
coordinate system in which the metric coefficients are already time
independent, then they will remain time independent during the
subsequent evolution, {\em i.e.} the lapse will not evolve and will
therefore not drive the time lines away from the Killing direction.
Second, the modified condition is naturally adapted to the use of a
densitized lapse as a fundamental variable, which in turn makes it a
good candidate for a dynamic slicing condition that can be used in
conjunction with some recently proposed hyperbolic reformulations of
the Einstein evolution equations.
\end{abstract}


\pacs{
04.20.Ex, 
04.25.Dm, 
95.30.Sf, 
}


\maketitle


\section{Introduction}
\label{sec:introduction}

Specifying a good foliation of spacetime is of fundamental importance
when studying the dynamical evolution of systems with strong
gravitational fields.  In the 3+1 decomposition of General Relativity,
a foliation of spacetime into spacelike hypersurfaces is described in
terms of the lapse function $\alpha$ that measures the interval of
proper time between neighboring hypersurfaces along their normal
direction.  The choice of a particular foliation represents the
freedom one has to specify the time coordinate and is therefore
arbitrary.  In practice, however, one can not choose a foliation ahead
of time since one is trying to solve for the geometry of the spacetime
itself, so one must choose instead a ``slicing condition'', that is,
some geometric or algebraic condition that allows one to calculate
the lapse function dynamically during the evolution.  Slicing
conditions come in many different forms, and they are usually chosen
by balancing their ease of implementation with the need to obtain a
well behaved coordinate system, as well as the well-posedness of the
resulting system of evolution equations.

In a recent paper~\cite{Alcubierre02b}, one of us studied a particular
slicing condition that has the property of being obtained through a
hyperbolic evolution equation for the lapse, and in many cases allows
one to construct a foliation that avoids different types of
pathological behaviors.  This slicing condition, known as the
Bona-Masso (BM) slicing condition~\cite{Bona94b}, has been used
successfully for many numerical simulations of strongly gravitating
systems such as black holes (see for
example~\cite{Arbona99,Alcubierre01a,Alcubierre02a}).  Other forms of
hyperbolic slicing conditions, which generalize in one way or another
the original proposal of Bona and Masso, have been suggested in the
last few years, examples of which can be found
in~\cite{Alcubierre02a,Sarbach02b}, and most recently
in~\cite{Lindblom03a}.

In this paper we argue that the original form of the BM condition (as
well as most of its generalization) has two important drawbacks:
First, it is not well adapted to the presence of Killing fields in the
sense that if one uses it to evolve a static spacetime, it can easily
drive the time lines away from the Killing direction.  Second, the
standard BM condition is not well adapted to the case when one wants
to use a densitized lapse as a fundamental variable.  Such a
densitized lapse has been recently advocated in the context of
hyperbolic reformulations of the Einstein equations (see for
example~\cite{Anderson99,Kidder01a}), and is therefore an important
issue to consider.  Here we will study a modified version of the BM
slicing condition that addresses both these issues at the same time.
This modified BM condition has already been used in the literature,
but as far as we are aware its properties have never been studied in
any detail.  On this manuscript we will limit ourselves to studying
the properties of this slicing condition independently of the Einstein
equations, and leave the study of how this condition couples to the
Einstein evolution equations for a separate work~\cite{Alcubierre03c}.

This paper is organized as follows.  In Sec.~\ref{sec:BM} we make a
brief introduction to the BM family of slicing conditions.
Section~\ref{sec:modified} motivates the introduction of a modified BM
condition from the point of view of compatibility with static and
stationary solutions.  We then study the relation of the modified
condition both with densitized lapses and with the divergence of the
coordinate time lines, and introduce a coordinate independent way of
writing the condition.  In Sec.~\ref{sec:hyperbolicity} we analyze the
hyperbolicity of the modified BM condition.  We start by discussing
the way in which the shift vector is chosen, and later analyze the
hyperbolicity of the coupled slicing-shift evolution system.  Finally,
Sections~\ref{sec:singularities} and~\ref{sec:shocks} study under
which circumstances the modified BM condition avoids focusing
singularities and gauge shocks. We conclude in
Sec.~\ref{sec:conclusion}.


\section{The Bona-Masso family of slicing conditions}
\label{sec:BM}

The BM family of slicing conditions~\cite{Bona94b} has been discussed
extensively in the literature (for a detailed discussion
see~\cite{Alcubierre00b} and references therein).  Here we will limit
ourselves to making a very brief description of its main properties.
The BM slicing condition is obtained by asking for the lapse function
to satisfy the following evolution equation
\begin{equation}
\frac{d}{dt} \; \alpha \equiv \left( \partial_t - {\cal L}_\beta \right)
\alpha = - \alpha^2 f(\alpha) \, K \\ ,
\label{eq:BonaMasso}
\end{equation}
with ${\cal L}_\beta$ the Lie derivative with respect to the shift
vector $\beta^i$, $K$ the trace of the extrinsic curvature and
$f(\alpha)$ a positive but otherwise arbitrary function of $\alpha$.
The condition above is a generalization of slicing conditions that
have been used in evolution codes based on the Arnowitt-Deser-Misner
(ADM) formulation~\cite{Arnowitt62,York79} since the early
90's~\cite{Bernstein93a,Anninos94c}, and was originally proposed in
the context of the Bona-Masso hyperbolic re-formulation of the Einstein
equations~\cite{Bona89,Bona92,Bona93,Bona94b,Bona97a}.  It is, however,
very general and can be used with any form of the 3+1 evolution
equations.  Since we will later introduce a modified version of the BM
slicing condition, we will refer from now on to
condition~(\ref{eq:BonaMasso}) as the ``standard BM condition''.

A very important property of the standard BM condition is the fact
that the shift terms included through the Lie derivative in
equation~(\ref{eq:BonaMasso}) are such that one is guaranteed to
obtain precisely the same spacetime foliation regardless of the value
of the shift vector.  At first sight, this would seem to be a natural
requirement for any slicing condition, but as we will see below this
might not be the most important property a slicing condition must
have.

The BM slicing condition can be shown to lead to a generalized wave
equation for the lapse
\begin{eqnarray}
&& \frac{d^2}{dt^2} \; \alpha - \alpha^2 f D^2 \alpha
= \nonumber \\
&& \hspace{3em} - \alpha^3 f \left[ K_{ij} K^{ij} - \left( 2 f
+ \alpha f' \right) K^2 \right] .
\label{eq:lapsewave}
\end{eqnarray}
From this equation one can easily see that the wave speed along a
specific direction $x^i$ given by
\begin{equation}
v_g = \alpha \sqrt{f \gamma^{ii}} \, ,
\label{eq:gaugespeed}
\end{equation}
where $\gamma_{ij}$ is the spatial metric.  The above expression
explains the need for $f(\alpha)$ to be positive: If it weren't, the
wave speed would not be real and the equation would be elliptic
instead of hyperbolic.  Notice also that the gauge
speed~(\ref{eq:gaugespeed}) can be smaller or larger that the speed of
light depending on the value of $f$.  Contrary to what one might
expect, having a gauge speed that is larger than the speed of light
does not lead to any causality violations, as the superluminal speed
is only related with the propagation of the coordinate system.  In
fact, empirically the most successful slicing conditions for the
simulation of black hole spacetimes have been precisely those that
allow superluminal gauge speeds (an example of this is the 1+log
slicing condition mentioned below).

Reference~\cite{Alcubierre02b} shows that, for some specific choices
of the function $f(\alpha)$, the standard BM slicing condition can
avoid both focusing singularities~\cite{Bona97a} and gauge
shocks~\cite{Alcubierre97a}.  We will expand on these points in the
following sections.

A very important particular case of the BM condition corresponds to
the choice $f(\alpha)=1$, which leads to the so-called ``harmonic
slicing'', for which the time coordinate obeys the simple wave
equation:
\begin{equation}
\Box t = g^{\mu \nu} \Gamma^0_{\mu \nu} = 0 \, ,
\label{eq:boxt}
\end{equation}
with $g_{\mu \nu}$ the spacetime metric. That is, $f=1$ corresponds to
the case when the time coordinate is a harmonic function.  One can
easily show that for harmonic slicing one also has the following
relation
\begin{equation}
\alpha = h(x^i) \; \gamma^{1/2}_n \, ,
\label{eq:harmonic}
\end{equation}
with $h(x^i)$ an arbitrary time independent function and
$\sqrt{\gamma_n}$ the volume element associated with observers moving
normal to the hypersurfaces (which will differ from the coordinate
volume element in the case of a non-vanishing shift vector).

Another particular case worth mentioning corresponds to ``1+log''
family, for which $f(\alpha)=N/\alpha$.  In this case we find the
following relation between the lapse function and the normal volume
elements
\begin{equation}
\alpha = h(x^i) + \ln \left( \gamma^{N/2}_n \right) \, ,
\end{equation}
which explains the name 1+log.  The case $N=2$ has been found
empirically to be particularly well
behaved~\cite{Arbona99,Alcubierre01a,Alcubierre02a}, something which
Ref.~\cite{Alcubierre02b} attributes (a posteriori) to the fact that
this is the only member of the 1+log family that avoids gauge shocks
even approximately.  Notice that for $N=2$ the gauge speed in
asymptotically flat regions where $\alpha \simeq 1$ becomes
\mbox{$\sqrt{2} > 1$}.  In regions inside a black hole, where the
lapse typically collapses to zero, the gauge speed can become
extremely large.

Having briefly discussed the main properties of the standard BM
slicing condition, we will now turn to our proposal for a modified
version of this condition, motivated by the analysis of stationary
spacetimes.


\section{A modified Bona-Masso slicing condition}
\label{sec:modified}

Let us consider for a moment a static or stationary spacetime, and let
us assume that we have chosen a coordinate system in which the metric
coefficients are already time independent.  Further, we will assume
that our coordinate system is such that the shift vector does not
vanish.  One would think at first sight that having a non-vanishing
shift implies that the spacetime can not be truly static but is at
most only stationary.  This is not entirely correct, as one can write
a static spacetime in a coordinate system with non-vanishing shift
and where the metric coefficients are nevertheless still time
independent.  An example of this is the Schwarzschild metric written
in Kerr-Schild or Painlev\'e-Gullstand coordinates.

Assuming we have such a situation, the ADM evolution equations imply
in particular that
\begin{eqnarray}
\partial_t \gamma_{ij} &=& - 2 \alpha K_{ij} + {\cal L}_\beta \gamma_{ij}
\nonumber \\
 &=&  - 2 \alpha K_{ij} + D_i \beta_j + D_j \beta_i = 0 \, ,
\end{eqnarray}
with $D_i$ the spatial covariant derivative.  From the last equation
one can obtain
\begin{equation}
\partial_t \gamma^{1/2} = - \gamma^{1/2} \left( \alpha K - D_i \beta^i
\right) = 0 \, ,
\label{eq:gammadot}
\end{equation}
with $\gamma^{1/2}$ the coordinate volume elements.  The trace of the
extrinsic curvature must therefore be given by
\begin{equation}
K = \frac{D_i \beta^i}{\alpha} \, .
\end{equation}

Let us now see if the standard Bona-Masso slicing condition is
compatible with the time-independent character of our spacetime in the
sense of predicting a time-independent lapse as well.  Substituting
our expression for $K$ into equation~(\ref{eq:BonaMasso}) we find
\begin{eqnarray}
\partial_t \alpha &=& \beta^i \partial_i \alpha - 2 \alpha^2 f(\alpha) K
\noindent \\
&=& \beta^i \partial_i \alpha - 2 \alpha f(\alpha) D_i \beta^i \, .
\end{eqnarray}
It is clear from looking at this expression that for arbitrary
$f(\alpha)$, $\partial_t \alpha$ will generally not vanish.  That is,
the lapse will evolve away from its preferred value and as a
consequence the spatial metric will not remain time independent.  This
indicates that the standard BM slicing condition is not well adapted
to the evolution of stationary spacetimes.

A modified version of the BM slicing condition that is well adapted to
such time independent spacetimes can nevertheless be easily obtained
by just asking for
\begin{equation}
\partial_t \alpha = \frac{\alpha f(\alpha)}{\gamma^{1/2}} \;
\partial_t \gamma^{1/2} \, ,
\label{eq:newBM0}
\end{equation}
which automatically guarantees that the lapse will not evolve if the
spatial metric is time independent.  One can integrate the last
equation trivially to find
\begin{equation}
\gamma^{1/2} = F(x^i) \, \exp \left\{ \int{\frac{d \alpha}
{\alpha f(\alpha)} } \right\} \; ,
\label{eq:gammaofalpha}
\end{equation}
with $F(x^i)$ a time independent function.  This tells us that there
is a very general functional relationship between $\alpha$ and
$\gamma^{1/2}$.  For $f(\alpha)=constant$ this relationship is a power
law, but in other cases it is more general (for example, the well
known ``1+log'' slicing is usually obtained from $f=2/\alpha$, which
gives us an exponential relation between $\alpha$ and $\gamma^{1/2}$).
Functional relationships between $\alpha$ and $\gamma^{1/2}$ have been
considered before in the context of finding hyperbolic re-formulations
of the 3+1 evolution equations.  For example, in
reference~\cite{Frittelli95}, Frittelli and Reula propose a general
power law relation, which as just mentioned is a particular form
of~(\ref{eq:gammaofalpha}) with $f(\alpha)$ constant. More recently,
Sarbach and Tiglio~\cite{Sarbach02b} have considered a completely
general functional relationship with the sole restriction that $d
\alpha / d \gamma > 0$, which clearly includes~(\ref{eq:gammaofalpha})
(asking for $f(\alpha)>0$ guarantees that $d \alpha / d \gamma >
0$).  What makes~(\ref{eq:gammaofalpha}) more interesting than a
very general relationship is the fact that one can learn a lot
about the properties of the slicing by studying the effect of
different forms of $f(\alpha)$.

Substituting now~(\ref{eq:gammadot}) into~(\ref{eq:newBM0}) we find
\begin{equation}
\partial_t \alpha = - \alpha f(\alpha) \left( \alpha K
- D_i \beta^i \right) \, .
\label{eq:newBM}
\end{equation}

This is the modified version of the BM slicing condition we wish to
study here (compare with~(\ref{eq:BonaMasso})). Several comments are
in order here.  First, we should mention that the modified BM
condition defined above has been used before in the literature in
Refs.~\cite{Alcubierre00a,Yo02a}.  Both these references study the
numerical evolution of a Schwarzschild spacetime written in
Kerr-Schild coordinates, and attempt to maintain the static (exact)
solution stable during the numerical simulation.  Because of this they
need to use a slicing condition that maintains the lapse equal to its
initial value in the continuum limit, but allows some dynamics to
respond to numerical truncation errors at finite resolutions. The
modified BM condition is used in those references precisely for the
reason outlined above, but no suggestion is made to use it in the more
general case or to study its properties.

Second, as condition~(\ref{eq:newBM}) does not include the Lie
derivative terms of the lapse with respect to the shift vector, it
{\em will not} give us the same foliation of spacetime for a different
choice of the shift vector, {\em i.e.} the foliation of spacetime one
obtains will depend on the choice of shift.  We believe this is not a
serious drawback since in a particular situation one would presumably
want to choose a slicing condition and a shift vector that are closely
interrelated.

Also, we believe that using a slicing condition that is compatible
with a static solution is a necessary requirement if one looks for
symmetry seeking coordinates of the type discussed by Gundlach and
Garfinkle~\cite{Garfinkle99a} and by Brady {\em et.al}~\cite{Brady98},
that will be able to find the Killing fields that static (or
stationary) spacetimes have, or the approximate Killing fields that
many interesting astrophysical systems will have at late times.  Of
course, having a slicing condition that is compatible with staticity
is not enough, one also needs to have a shift condition that has the
same property.  Otherwise, the shift evolution will also drive us away
from the frame in which the staticity is apparent.  In this paper we
will not deal with the issue of the shift choice, but will consider it
in a future work.

And finally, even if one has gauge conditions (for both lapse and shift)
that keep a static solution static, this does not guarantee that one
can find the Killing direction if one starts in the wrong coordinate
system. All that it guarantees is that if we do find this Killing
direction we won't be driven away from it again.


\subsection{Densitized lapse}
\label{sec:densitized}

Since the early 90's, many re-formulations of the 3+1 evolution
equations of general relativity have been
proposed~\cite{Bona92,Bona93,
Bona94b,Choquet95,Frittelli95,Frittelli99,Friedrich96,vanPutten95,
Abrahams96a,Bona97a,Abrahams97b,Anderson97,Anderson99,Alcubierre99c,
Kidder01a}.  The main purpose of most such re-formulations has been to
recast the Einstein evolution equations as a strongly or symmetric
hyperbolic system motivated by the fact that for such systems one can
prove that the evolution equations are well posed, and with the hope
that such well-posedness will bring with it improvements in numerical
simulations, both in terms of stability and in terms of allowing one
to impose boundary conditions in a consistent and well behaved way.

While it is not the purpose of this paper to study the different
hyperbolic formulations of the Einstein evolution equations, there is
one related point that is crucial for the discussion of slicing
conditions.  It was realized early on that there were two main
problems when trying to go from the standard ADM evolution equations
to a strongly or symmetric hyperbolic formulation.  One problem was
associated with the existence of constraint violating solutions that
spoil hyperbolicity and required for its solution the addition of
multiples of the constraint equations, in a number of different ways,
to the evolution equations (the constraint violating modes are not
eliminated, just transformed in a way that allows a hyperbolic system
to be constructed).

The second problem, more directly related to our discussion, was the
observation that it is not possible to construct a strongly hyperbolic
formulation of the 3+1 evolution equations if the lapse function is
considered to be an a priori known function of space and time.  Two
different routes have been followed to solve this second problem.  The
first route, taken in the Bona-Masso hyperbolic
re-formulation~\cite{Bona92,Bona93,Bona94b,Bona97a}, was to propose an
evolution equation for the lapse (equation~(\ref{eq:BonaMasso}) of the
previous section), and then construct a strongly hyperbolic system of
equations where the lapse was considered just another dynamical
variable.  This same route has been followed very recently by Lindblom
and Scheel~\cite{Lindblom03a}, where generalizations of standard BM
slicing condition and the ``$\Gamma$-driver'' shift
condition~\cite{Alcubierre02a} have been used to construct a symmetric
hyperbolic system that includes lapse and shift as dynamical
variables.  The second route has been to take not the lapse, but
rather the {\em densitized} lapse $q:=\alpha \gamma^{-1/2}$, to be a
prescribed function of space and time (see~\cite{Anderson99,Kidder01a}
and references therein).  Both these routes have been successful in
constructing strongly hyperbolic re-formulations of the Einstein
evolution equations.  The different approaches are related, but are
not equivalent, as can be seen easily if we consider for a moment the
BM slicing condition in the case $f=1$ (the harmonic slicing case).
As we have seen, in that case the lapse takes the form
\begin{equation}
\alpha = h(x^i) \; \gamma^{1/2}_n \, ,
\end{equation}
with $\gamma^{1/2}_n$ the volume elements associated with the normal
observers, and $h(x^i)$ an arbitrary function of space.  We can in
fact turn the function $h(x^i)$ into an arbitrary function of both
space {\em and} time if we add a source term to the standard BM
condition in the following way
\begin{equation}
\left( \partial_t - {\cal L}_\beta \right)
\alpha = - \alpha^2 f(\alpha) \, K + H(x,t) \ ,
\end{equation}
with $H(x,t)$ arbitrary.  On the other hand, the condition for the
densitized lapse to be a known function of spacetime takes the form
\begin{equation}
\alpha = q(x,t) \; \gamma^{1/2} \, ,
\end{equation}
where $\gamma^{1/2}$ are the {\em coordinate} volume elements.  In the
case of a vanishing shift vector, normal and coordinate volume
elements coincide, and the BM condition can be seen as a
generalization of the prescribed densitized lapse condition.  But for
non-zero shift vector this is no longer the case.

One can in fact rewrite the standard BM condition in terms of the
densitized lapse $q$ in the following way
\begin{equation}
\left( \partial_t - {\cal L}_\beta \right) \; q = - q^2 \gamma^{1/2}
\left( f - 1 \right) K  \, ,
\end{equation}
which shows that even for $f=1$, the densitized lapse $q$ will evolve
dynamically driven by the Lie derivative term.

The crucial observation here is that the modified BM
condition~(\ref{eq:newBM}), when written in terms of the densitized
lapse, takes the form
\begin{equation}
\partial_t q = - q \left( f - 1 \right) \left[ q \; \gamma^{1/2} K
- D_i \beta^i \right] \, .
\label{eq:densitized}
\end{equation}
It is now clear that by taking $f=1$ this equation reduces to the case
of a static densitized lapse (the case of a prescribed densitized
lapse that is not time independent can be easily obtained by adding a
source term to the above equation).  The modified BM slicing condition
would therefore seem to be a natural generalization of the prescribed
densitized lapse, and should be well adapted to hyperbolic
re-formulations of the Einstein equations that use of a densitized
lapse~\cite{Alcubierre03c}.


\subsection{Divergence of the time lines and Killing fields}
\label{sec:divergence}

From the ADM equations one can easily show that the divergence of the
time lines is given by the following expression
\begin{equation}
\nabla_\mu t^\mu = \frac{1}{\alpha} \left[ \partial_t \alpha
- \alpha \left( \alpha K - D_i \beta^i \right) \right] \; ,
\label{eq:divergence1}
\end{equation}
where $t^\mu$ is the vector tangent to the 3+1 time lines, which is
defined in terms of the normal and shift vectors as
\begin{equation}
t^\mu = \alpha n^\mu + \beta^\mu \; .
\label{eq:timevector}
\end{equation}

Equation~(\ref{eq:divergence1}) implies that the modified BM slicing
condition can also be written as
\begin{equation}
\partial_t \alpha = \alpha \left( \frac{f}{f+1} \right) \nabla_\mu t^\mu \; .
\label{eq:divergence2}
\end{equation}

The last equation shows that the evolution of the lapse is directly
related to the divergence of the time lines.  We then see that just as
the original BM slicing condition was such that the lapse reacted to
the divergence of the normal observers, the modified condition ensures
that the lapse reacts to the divergence of the coordinate time lines.

Let us assume for the moment that our spacetime has a future-pointing
Killing field $v^\mu$.  In that case we will have
\begin{equation}
\nabla_\mu v_\nu + \nabla_\nu v_\mu = 0 \qquad \Rightarrow \qquad
\nabla_\mu v^\mu = 0 \; .
\label{eq:killing}
\end{equation}
If we now assume that our time lines are oriented along the Killing
direction, then equation~(\ref{eq:divergence2}) automatically implies
that the lapse function will be time independent.  This, of course, we
already knew since it was our initial motivation for modifying the BM
condition.


\subsection{Generalized wave equation for the time function}
\label{sec:foliation}

In reference~\cite{Alcubierre02b} it was shown that the standard
BM slicing condition can be written as a generalized wave equation
for a ``time function'' $\phi$ in the following way
\begin{equation}
\left( g^{\mu \nu} - a \; n^\mu n^\nu \right)
\nabla_\mu \nabla_\nu \phi  = 0 \; ,
\label{eq:foliationBM}
\end{equation}
with $n^\mu$ the unit normal vector to the spatial hypersurfaces and
$a := 1/f(\alpha) -1$.  The different members of the spacetime
foliation can then be obtained as the level sets of the time function
$\phi$.

One can also construct such a foliation equation for the modified BM
condition~(\ref{eq:newBM}).  The corresponding equation for the time
function $\phi$ turns out to be
\begin{equation}
\left( g^{\mu\nu} - \frac{a}{\alpha}\,t^\mu\,n^\nu \right)
\nabla_\mu \nabla_\nu \,\phi
= \nabla_\mu\,\left( \frac{\beta^\mu}{\alpha^2} \right) \; .
\label{eq:foliation}
\end{equation}

By writing equation~(\ref{eq:foliation}) in the standard 3+1
coordinate system adapted to the foliation, it is not difficult to
show that it is in fact equivalent to the slicing
condition~(\ref{eq:newBM}).  On the other hand, notice that when
written in a different coordinate system, the vector $t^\mu$ does not
have to be aligned with the time lines any longer.  Also, the shift
vector $\beta^\mu$ will have a non-zero time component.  However, we
will still have
\begin{equation}
\beta^\mu n_\mu = 0 \; .
\label{eq:parallelshift}
\end{equation}

In the case of vanishing shift, equation~(\ref{eq:foliation}) reduces
to equation~(\ref{eq:foliationBM}), which is just another way of
saying that in that case the original and modified BM conditions
coincide.  For non-vanishing shift, however, both equations differ.
The foliation equation for the original BM condition makes no reference
to the shift, which implies that the foliation is independent of our
choice of shift (as already mentioned above).  The modified foliation
equation, however, clearly depends on the shift choice, so the
foliation of spacetime one obtains will depend on the shift as well.

The foliation equation~(\ref{eq:foliation}) is very useful when trying
to understand the properties of the slicing condition in a covariant
way that is independent of the Einstein field equations.


\section{Hyperbolicity}
\label{sec:hyperbolicity}

The concept of hyperbolicity is of fundamental importance in the study
of the evolution equations associated with a Cauchy problem.  Some
measure of hyperbolicity, even in a weak sense, implies that the
system of equations is causal, {\em i.e.} that the solution at a given
point in spacetime depends only on data in a region of compact support
to the past of that point (the characteristic cone).  Stronger
versions of hyperbolicity can also be used to prove rigorously that
the system of equations is well-posed, that is, that its solutions
exist (at least locally), are unique, and are stable in the sense that
small changes in the initial data will correspond to small changes in
the solution.  Hyperbolicity also allows one to construct well-posed
initial-boundary problems, which implies that one should be able to
obtain well behaved boundary conditions for numerical simulations with
artificial boundaries.

Because of this, showing that a given system of evolution equations is
hyperbolic has become an important test for new formulations of the
3+1 equations.  In our case, since we are studying a slicing condition
that is obtained through an evolution equation for the lapse, the
question of the hyperbolicity of the gauge condition becomes
important.  Since we want to look at this issue in a way that is
independent of the Einstein equations, we will consider from now on a
given background spacetime (which may or may not obey Einstein's
equations), and study our slicing condition on this fixed background.


\subsection{Prescribed shift vector}
\label{sec:shift}

Since the foliation equation~(\ref{eq:foliation}) involves the shift
vector $\beta^\mu$, in order to analyze its hyperbolicity we must say
something about how the shift vector evolves in time.  The simplest
approach is to assume that we have a prescribed, {\em i.e.}
non-dynamical, shift vector.  However, this immediately leads us into
the question: What does it mean to have a prescribed shift vector in a
covariant sense?

Clearly, we can not just assume that the shift vector is an {\em a
priori} known vector-field in spacetime.  This is because the shift
vector must always be parallel to the spatial hypersurfaces making up
our foliation, and those hypersurfaces are precisely what we are
trying to solve for.  This means that inevitably, as we solve for the
hypersurfaces, the shift vector must evolve dynamically to guarantee
that it remains parallel to them.  At most, we can ask for the
magnitude and direction of the shift vector {\em within a given
hypersurface} to be prescribed functions of space and time.

We will then propose the most general evolution equation for the shift
that is compatible with the fact that the shift lives on the spatial
hypersurfaces.  For this we proceed as follows: On a given
hypersurface we can choose a basis of spatial vectors $e^\mu_i$, and
express the shift vector in terms of such a basis:
\begin{equation}
\beta^\mu = e^\mu_i \, b^i \; ,
\end{equation}
where the $b^i$ are the components of the shift vector in the basis
under consideration. We will now identify this basis with the standard
3+1 spatial coordinate basis.  This means that the spatial basis
$e^\mu_i$, together with the time vector $t^\mu$, form a coordinate
basis for the spacetime at the point under study (they form the
standard 3+1 coordinate basis), which implies the following commutation
relation
\begin{equation}
t^\nu\,\partial_\nu\,e^\mu_i = e^\nu_i\,\partial_\nu\,t^\mu \; .
\end{equation}

Using this relation, we can obtain the following very general
evolution equation for the shift vector:
\begin{equation}
t^\nu \, \partial_\nu \, \beta^\mu
= \beta^\nu \, \partial_\nu \, t^\mu + s^\mu \; ,
\label{eq:shift}
\end{equation}
where $s^\mu=e^\mu_i \, t^\nu \, \partial_\nu\,b^i$.  Notice that the
last equation is fully covariant even if it is written in terms of
partial derivatives, as the Christoffel symbols cancel out.  If we now
assume that the components $b^i$ of the shift in the 3+1 spatial
coordinate basis are prescribed functions of spacetime, then we can
consider the $s^\mu$ as source terms in the hyperbolicity analysis.

We will use equation~(\ref{eq:shift}) above as our general evolution
equation for the shift.  Notice that in the particular case when we
restrict ourselves to the 3+1 coordinate system, for which $t^\mu =
(1,0,0,0)$, this equation simply reduces to
\begin{equation}
\partial_t \beta^i = \partial_t b^i \; ,
\end{equation}
 which is of course to be expected.


\subsection{Hyperbolicity of the foliation equation}

We are interested in studying the hyperbolicity of the modified BM
slicing condition in a way that is independent of the Einstein
equation.  In order to do this we assume we have a fixed background
spacetime with metric $g_{\mu \nu}$, and study the foliation
equation~(\ref{eq:foliation}), which we repeat here for clarity:
\begin{equation}
\left( g^{\mu\nu} - \frac{a}{\alpha} \, t^\mu \, n^\nu \right)
\nabla_\mu \nabla_\nu \, \phi
= \nabla_\mu\,\left( \frac{\beta^\mu}{\alpha^2} \right) \; .
\label{eq:phi}
\end{equation}
Here $\phi$ is a scalar function whose level sets identify the
elements of the foliation, and $a := 1/f(\alpha)-1$.  In terms of
$\phi$, the lapse function and the unit normal vector can be expressed
as
\begin{eqnarray}
\alpha &=& \left( - \nabla_\mu \, \phi \;
\nabla^\mu \, \phi \right)^{-1/2} \; ,
\label{eq:alpha} \\
n_\mu &=& -\alpha \nabla_\mu \, \phi \; .
\label{eq:n}
\end{eqnarray}

We will now concentrate on a point on a given slice, and construct
locally flat coordinates in its neighborhood.  We will further define
the following first order quantities
\begin{eqnarray}
\partial_t \phi & \equiv & \Pi \; , \\
\partial_i \phi & \equiv & \Psi_i \; .
\end{eqnarray}
The lapse and unit normal vector then become
\begin{eqnarray}
\alpha &=& \left(\Pi^2-\Psi^2 \right)^{-1/2} \; , \\
n^\mu &=& \alpha\, (\Pi,-\Psi^i) \; ,
\end{eqnarray}
with $\Psi^2 \equiv \sum \Psi_i^2$. Notice that, as we are in locally
flat coordinates, there is no difference between lower and upper
spatial indexes, so we will be using them indiscriminately.

On the other hand, since the shift vector is parallel to the
hypersurface we must have
\begin{equation}
\beta^\mu n_\mu = 0 \; ,
\end{equation}
which allows us to express the $\beta^0$ component as
\begin{equation}
\beta^0 = -\frac{\beta^i \Psi_i}{\Pi} \; .
\end{equation}

With the definitions above we can now rewrite the foliation
equation~(\ref{eq:phi}) as:
\begin{eqnarray}
- \left( P \,\alpha^2 - \left( (1+a)\Pi^2+\Psi^2 \right)
\,Q \right) \partial_t \Pi
+ \frac{\Psi_i}{\alpha^2\Pi} \, \partial_t \beta^i
 && \nonumber \\
+ B^m \partial_m \Pi + C^{mj} \partial_m \Psi_j -
\frac{1}{\alpha^2} \; \partial_m \beta^m = 0 \; , \hspace{5mm} &&
\label{eq:d1}
\end{eqnarray}
where dot stands for the partial time derivative, and where we have defined
\begin{eqnarray}
P &:=& (1+a) \, \Pi^2 - \Psi^2 \; , \\
Q     &:=& \frac{\beta^i \Psi_i}{\Pi^2} \; , \\
B^m   &:=& -\left( (1+a) \Pi^2 + \Psi^2 \right) \frac{\beta^m}{\Pi}
\nonumber \\
&+& \left( 2 a \alpha^2 - \left( 2 + a \right) Q \right) \Pi \Psi^m , \\
C^{mj}&:=& \delta^{mj} - \Psi^m  \left(a \alpha^2 \Psi^j -
(2+a) \beta^j \right) \; .
\end{eqnarray}
Notice that to arrive at Eq.~(\ref{eq:d1}) above we have used the fact
that
\begin{equation}
\partial_t \Psi_i = \partial_i \Pi \; ,
\label{eq:psidot}
\end{equation}

We can also rewrite the evolution equation for the shift introduced in
the last section, Eq.~(\ref{eq:shift}), in our locally flat
coordinates to find
\begin{eqnarray}
2 \alpha^2 \Psi_m \beta^m \Psi^i \partial_t \Pi + \Pi \partial_t \beta^i -
N^{im} \partial_m \Pi && \nonumber \\
+ L^{ijm} \partial_m \Psi_j - \Psi^m \partial_m \beta^i
- \frac{s^i}{\alpha^2} &=& 0 \; , \hspace{5mm}
\label{eq:d2}
\end{eqnarray}
where now
\begin{eqnarray}
N^{im} &:=& \Pi Q \delta^{im} +
2 \alpha^2 \Pi \Psi^i \left( Q \Psi^m + \beta^m \right) , \\
L^{ijm} &:=& \left( \delta^{ij} + 2 \alpha^2 \Psi^i \Psi^j
\right) \beta^m \; .
\end{eqnarray}

Equations (\ref{eq:d1}) and (\ref{eq:d2}) can now be used to find the
following evolution equations for $\Pi$, and the spatial part of the
shift vector $\beta^i$:
\begin{eqnarray}
\partial_t \Pi &=& \frac{1}{T_1} \left( \rule{0mm}{4mm} T^{m}_2 \partial_m \Pi
 \right. \nonumber \\
&+& \left. T^{jm}_3 \partial_m \Psi_j + T^{jm}_4 \partial_m \beta^j +
F^{0} \right) \; , \label{eq:pidot} \\
\partial_t \beta^i &=& D^{im}_1 \partial_m \Pi +
D^{ijm}_2 \partial_m \Psi_j \nonumber \\
&+& D^{ijm}_3 \partial_m \beta^j + F^i \; , \label{eq:betadot}
\end{eqnarray}
where
\begin{eqnarray}
T_1 &=& P (Q-\alpha^2),\\
T^m_2 &=& \frac{P}{\Pi}\,\beta^m-\Pi\,\Psi^m\,\left( 2a\alpha^2
-\frac{PQ}{\Pi^2}\right),\\
T^{jm}_3&=&-\delta^{jm}+ a\,\Psi^j\,(\alpha^2\,\Psi^m-\beta^m) -
\frac{\beta^j\,\Psi^m}{\alpha^2\,\Pi^2},\\
T^{jm}_4&=&\frac{1}{\alpha^2}\left(
\delta^{jm}-\frac{\Psi^j\,\Psi^m}{\Pi^2}\right),\\
F^0 &=& \frac{\Psi_i\,s^i}{\alpha^4\,\Pi},\\
D^{im}_1 &=& Q \left[ \delta^{im} +
2 \alpha^2 \Psi^i \left( \Psi^m + \frac{\beta^m}{Q} -
\Pi \frac{T^m_2}{T_1} \right) \right] , \\
D^{ijm}_2 &=& - \frac{\delta^{ij} \beta^m}{\Pi}
- \frac{2 \alpha^2 \Psi^i}{\Pi} \left( \Psi^j \beta^m
+ \frac{Q \Pi^2 T^{jm}_3}{T_1} \right) , \hspace{8mm} \\
D^{ijm}_3 &=& \frac{1}{\Pi} \left( \delta^{ij} \Psi^m -
2 \alpha^2 \Pi^2 Q \Psi^i \frac{T^{jm}_4}{T_1} \right),\\
F^i &=& \frac{1}{\Pi} \left( \frac{s^i}{\alpha^2} -
2 \alpha^2 \Pi^2 Q \Psi^i \frac{F^0}{T_1} \right).
\end{eqnarray}

Equations~(\ref{eq:pidot}) and (\ref{eq:betadot}), together with
Eq.~(\ref{eq:psidot}), form our closed set of evolution equations.
Notice that these equations are only valid if $\alpha^2 \neq Q$, the
case \mbox{$\alpha^2 = Q$} being degenerate.  The reason for this can
be traced back to the fact that if $\alpha^2=Q$, then $\beta^i \Psi_i
= (\alpha \Pi)^2$, which in implies that $t^0 = \alpha n^0 + \beta^0 =
\alpha^2 \Pi - \beta^i \Psi_i / \Pi = 0$.  This means that the vector
$t^\mu$ has no time component and equation~(\ref{eq:shift}) stops
being an evolution equation for the shift.  That this is a purely
coordinate problem can be seen from the fact that we can always boost
our locally flat coordinates in such a way that $\Psi_i$ becomes zero
and the problem disappears.

In order to determine whether or not the system of evolution equations
is hyperbolic, we first write it in matrix notation. We start by
defining the vector ${\bf u}$ in the following way
\begin{equation}
{\bf u} =( \Pi,\Psi_x,\Psi_y,\Psi_z,\beta^x,\beta^y,\beta^z ) \; .
\label{eq:variables}
\end{equation}
Thus the system of equations can be expressed as
\begin{equation}
\partial_t \, {\bf u} = {\bf M}^x \, \partial_x \, {\bf u}
+ {\bf M}^y \, \partial_y \, {\bf u} + {\bf M}^z \, \partial_z\,{\bf u}
+ {\bf s} \; ,
\label{eq:system}
\end{equation}
where the Jacobian matrices ${\bf M}^i$ and the source vector ${\bf
s}$ depend on the $u$'s but not their derivatives.

The matrix ${\bf M}^x$ has the particular form
\begin{equation}
\bf{M}^x =
\left(
\begin{array}{c c c}
T_2^x/ T1 & T_3^{jx} / T1 & T_4^{jx} / T1 \\
1 & 0 & 0 \\
0 & 0 & 0 \\
0 & 0 & 0 \\
D_1^{ix} & D_2^{ijx} & D_3^{ijx}
\end{array} \right) \; ,
\end{equation}
with $i,j=x,y,z$.  The matrices ${\bf M}^y$ and ${\bf M}^z$ have
similar structures.

Having written our system of equations in the form~(\ref{eq:system}),
we can now proceed to study its hyperbolicity properties.  In order to
do this one should first construct the ``principal symbol'' $M^i n_i$,
where $n_i$ is an arbitrary unit vector.  The system is then said to
be strongly hyperbolic if the eigenvalues of the principal symbol are
real and there is a complete set of eigenvectors for all $n_i$;
furthermore, the system is said to be symmetric hyperbolic if the
principal symbol can be symmetrized in a way that is independent of
the $n_i$~\cite{kreiss89}.  If, on the other hand, the eigenvalues are
real, but there is no complete set of eigenvectors, the system is only
weakly hyperbolic.  Here we will use a shortcut useful when all
directions have equivalent structures: We will find the eigenvalues of
one of the Jacobian matrices $M^i$ which by inspection can
later be generalized to any arbitrary direction.  We will concentrate
on the matrix $M^x$, as results for the other two matrices can be
obtained afterward in a straightforward way.

The eigenvalues of ${\bf M}^x$ can be found to be
\begin{eqnarray}
\lambda_{\pm} &=& \frac{\Pi}{P} \left( a \Psi_x \pm R \right) \; , \\
\lambda^{(3)} &=& \frac{\beta^x - \Psi_x \alpha^2}{\Pi (Q - \alpha^2)} \; ,\\
\lambda^{(4)} &=& \lambda^{(5)} = \frac{\Psi_x}{\Pi} \; , \\
\lambda^{(6)} &=& \lambda^{(7)} = 0 \; ,
\end{eqnarray}
with $R=\sqrt{P - a\,\Psi_x^2} / \alpha \Pi$.

As expected, two of the eigenvalues are zero due to the fact that two
rows of the matrix are zero.  Of the remaining eigenvalues,
$\lambda^{(3)}$, $\lambda^{(4)}$ and $\lambda^{(5)}$ are clearly real,
but $\lambda_{\pm}$ involve square roots through $R$ so they require a
more careful analysis.  However, by inspecting the expressions for
$\lambda_{\pm}$ it is easy to see that, though written in a different
way, they are in fact identical to the eigenvalues found in
Ref.~\cite{Alcubierre02b} for the case of the standard BM slicing
condition.  Since in that reference it was shown that those
eigenvalues are always complex for $f(\alpha)<0$, always real for
$0 <f(\alpha) \leq 1$ and can always be made real for $f(\alpha)>1$ by
an adequate orientation of the coordinate system, we conclude that our
system of equations of equations is hyperbolic, at least weakly, as
long as $f>0$.

The fact that two of the eigenvalues turn out to be identical to those
found in the case of the standard BM slicing condition is surprising.
One could have expected that, since the modified BM condition depends
on the choice of shift, the shift vector should have affected these
eigenvalues.  The fact that it leaves these eigenvalues unchanged can
be more easily understood in the 3+1 coordinate frame.  We carry out
this analysis in the Appendix.

Having found that our equations are at least weakly hyperbolic, we now
want to show that they are in fact also strongly hyperbolic. For this
we must see if we have a complete set of eigenvectors.  The
eigenvectors associated with the matrix ${\bf M}^x$ are:
\begin{eqnarray}
{\bf e}_{\pm} &=& [\lambda_{\pm},1,0,0,(\Psi_x\,B_\pm \pm
R)\,A_\pm, \nonumber \\
&&\Psi_y\,B_\pm\,A_\pm,\Psi_z\,B_\pm\,A_\pm ] \; , \\
{\bf e}^{(3)} &=& [\frac{1}{\alpha^2} \left( \Psi_x -
\frac{\beta^x}{\alpha^2} \right), \frac{\Pi}{\alpha^2}\left(1 -
\frac{Q}{\alpha^2} \right),0,0, \nonumber \\
&& \Pi \, \left( 1 - \frac{Q}{\alpha^2} \right) +
\Psi_x\,W, \Psi_y\,W,\Psi_z\,W ] \; , \hspace{6mm} \\
{\bf e}^{(4)} &=& [0,0,0,0,\frac{\Psi_x \Psi_y}{\Pi^2 - \Psi_x^2},1,0 ] \; , \\
{\bf e}^{(5)} &=& [0,0,0,0,\frac{\Psi_x \Psi_z}{\Pi^2 - \Psi_x^2},0,1 ] \; , \\
{\bf e}^{(6)} &=& [0,Z^1,1,0,Z^2,Z^3,Z^4] \; , \\
{\bf e}^{(7)} &=& [0,Z^5,0,1,Z^6,Z^7,Z^8] \; ,
 \label{eq:eigve}
\end{eqnarray}
where
\begin{eqnarray}
B_\pm &=& 2 \left( \frac{\Psi_x}{\Pi} \, \lambda_\pm - 1 \right) \; , \\
A_\pm &=& \frac{\alpha^2 \Pi^2}{(\Pi^2 - \Psi_x^2)(P - a \Psi_x^2)}
\left\{ ( \alpha^{-2} - a \Psi_x^2 )
Q \Psi_x \rule{0mm}{5mm}  \right. \nonumber \\
&- & \left. ( P - 2 a \Psi_x^2 ) \beta^x +
( \beta^y \Psi_y + \beta^z \Psi_z ) \,
\frac{P\lambda_\pm}{\Pi} \right\} , \hspace{8mm} \\
W &=& 2\Pi\left( \beta^x - Q \Psi_x \right) \; .
\end{eqnarray}
The functions $Z^i$ are somewhat long expressions whose explicit form
is not needed in what follows, so we will not write them here.

It is well known that those eigenvectors corresponding to different
eigenvalues are linearly independent. The eigenvectors associated with
the degenerate eigenvalues $\lambda^{(4)},\lambda^{(5)}$ and
$\lambda^{(6)},\lambda^{(7)}$, can be seen to be independent from one
another by inspection. Therefore, the eigenvectors given above form a
complete set. This, together with the fact that the eigenvalues are
real for $f>0$ allows us to conclude that the system of evolution
equations~(\ref{eq:psidot},\ref{eq:pidot},\ref{eq:betadot}) is
strongly hyperbolic if $f>0$.


\section{Singularity avoidance}
\label{sec:singularities}

In reference~\cite{Alcubierre02b} it was shown that the original BM
slicing condition can avoid so-called ``focusing
singularities''~\cite{Bona97a} depending on the form that the function
$f(\alpha)$ takes in the limit of small $\alpha$.  In particular, it
was shown that if $f(\alpha)$ behaves as~$f=A \alpha^n$ for small
$\alpha$ and the normal volume elements vanish in terms of proper time
$\tau$ as $\gamma_n^{1/2} \sim \left( \tau_s - \tau \right)^m$, one
can have three different types of behavior:

\begin{enumerate}

\item For $n<0$ the lapse vanishes before the normal spatial volume
elements do, which corresponds to strong singularity avoidance

\item For $n=0$ and $m A \geq 1$ the lapse vanishes with the normal
spatial volume elements and the singularity is reached after an
infinite coordinate time, corresponding to marginal singularity
avoidance

\item For both $n>0$, and $n=0$ with $m A < 1$, the lapse vanishes
with the normal volume elements but the singularity is still reached
in a finite coordinate time, so there is no singularity avoidance.

\end{enumerate}

The results summarized above, however, depended crucially on the fact
that the original BM slicing condition relates the evolution of the
lapse to the evolution of the normal volume elements.  The modified
version of the BM condition, on the other hand, relates the evolution
of the lapse to the evolution of the {\em coordinate}\/ volume elements.
We must therefore see how this affects the conclusions about
singularity avoidance.  The first thing to notice is that all the
analysis of reference~\cite{Alcubierre02b} will follow exactly in the
same way for the modified BM condition if we replace normal volume
elements with coordinate volume elements.  This means that the
modified BM condition will avoid singularities where the coordinate
volume elements vanish ({\em i.e.} coordinate focusing singularities)
under the same conditions as before.

There is one very important difference between ``normal'' focusing
singularities and ``coordinate'' focusing singularities.  When the
normal volume elements vanish, the normal direction to the
hypersurfaces becomes ill-defined and the hypersurfaces become
non-smooth.  When the coordinate volume elements vanish, on the other
hand, it is only the time lines that cross.  This means that one could
in principle develop a coordinate focusing singularity on a perfectly
smooth hypersurface, or worse still, one could develop a normal
focusing singularity for which the time lines do not cross and the
coordinate volume elements remain non-zero.  The second case would be
extremely problematic as our hypersurfaces would become non-smooth but
the lapse would not collapse in response to this.

To see under what conditions we can have one type of focusing
singularity and not the other we must look at the evolution equations
for the normal and coordinate volume elements:
\begin{eqnarray}
\partial_t \ln \gamma_n^{1/2} &=& - \alpha K \; , \\
\partial_t \ln \gamma^{1/2} &=& - \left( \alpha K - D_i \beta^i
\right) \; .
\end{eqnarray}
From the first of these equations it is clear that for a normal
focusing singularity to develop \mbox($\gamma_n^{1/2} \rightarrow 0$)
we must have \mbox{$K \rightarrow \infty$}.  From the second equation
we then see that the only way in which we can have a normal focusing
singularity develop while at the same time keeping a non-zero
coordinate volume element is for $D_i \beta^i$ to diverge with $K$ while
keeping their difference finite.

We then conclude that if the divergence of the shift remains finite,
both types of focusing singularities will happen at the same time.
This means that if the shift vector remains regular, the modified BM
slicing condition will avoid singularities exactly in the same way in
which the original condition did.


\section{Gauge shocks}
\label{sec:shocks}

In Section~\ref{sec:hyperbolicity} we have shown that the system of
equations (\ref{eq:system}) for the variables~(\ref{eq:variables}) is
strongly hyperbolic.  We can then define a complete set of
``eigenfields'' \mbox{$\omega_i$} in the following way:
\begin{equation}
{\bf u} = R \, {\bf \omega} \qquad \Rightarrow \qquad 
{\bf \omega} = R^{-1}  \, {\bf u}  \; ,
\end{equation}
where \mbox{$R$} is the matrix of column eigenvectors \mbox{$\bf{e}_i$}.

We say that the eigenfield \mbox{$\omega_i$} is ``linearly degenerate'' if
its corresponding eigenvalue \mbox{$\lambda_i$} is independent of the
eigenfield, that is
\begin{equation}
\frac{\partial \lambda_i}{\partial \omega_i} = 
\sum^{N_u}_{j=1} \, \frac{\partial \lambda_i}{\partial u_j} \,
\frac{\partial u_j}{\partial \omega_i} = 
\nabla_u \lambda_i \cdot {\bf e}_i = 0 \; .
\label{eq:lindeg}
\end{equation}
Linear degeneracy guarantees that the corresponding eigenfield will
not develop shocks.

Using the eigenvalues and eigenvectors found in
Sec.~\ref{sec:hyperbolicity}, the conditions for linear degeneracy
become
\begin{eqnarray}
C_{\pm}  &:=& \lambda_{\pm} \, \frac{\partial \lambda_{\pm}}{\partial \Pi} 
+ \frac{\partial \lambda_{\pm}}{\partial \Psi_{x}} = 0 \; , \label{eq:cpm} \\
C_{3}  &:=& {\bf e}^{(3)}_{1} \, \frac{\partial \lambda^{(3)}}{\partial \Pi} 
+ {\bf e}^{(3)}_{2} \, \frac{\partial \lambda^{(3)}}{\partial \Psi_{x}}
+ {\bf e}^{(3)}_{5} \, \frac{\partial \lambda^{(3)}}{\partial \beta^{x}}
\nonumber \\
&+& {\bf e}^{(3)}_{6} \, \frac{\partial \lambda^{(3)}}{\partial \beta^{y}}
+ {\bf e}^{(3)}_{7} \, \frac{\partial \lambda^{(3)}}
{\partial \beta^{z}} = 0 \; , \label{eq:c3}  \\
C_{4} &:=& {\bf e}^{(4)}_{5} \,
\frac{\partial \lambda^{(4)}}{\partial \beta^{x}} 
+ \frac{\partial \lambda^{(4)}}{\partial \beta^{y}} = 0 \; , \label{eq:c4} \\
C_{5} &:=& {\bf e}^{(5)}_{5} \, \frac{\partial \lambda^{(5)}}
{\partial \beta^{x}}  + \frac{\partial \lambda^{(5)}}
{\partial \beta^{z}} = 0 \; , \label{eq:c5} \\
C_{6} &=& C_{7} = 0 \label{eq:c67} \; .
\end{eqnarray}

A straightforward calculation shows that equations
(\ref{eq:c3})-(\ref{eq:c5}) are satisfied identically.  On the other
hand, equation~(\ref{eq:cpm}) is precisely the same equation found for
the standard BM slicing condition in reference~\cite{Alcubierre02a},
where it was shown that it leads to the following condition on the
function $f(\alpha)$
\begin{equation}
1-f-\frac{\alpha f^{\prime}}{2} = 0 \; .
\end{equation}
This means that the modified BM slicing condition is linearly
degenerate under the same circumstances as the standard BM condition,
and therefore it will avoid gauge shocks exactly in the same cases.


\section{Conclusion}
\label{sec:conclusion}

We have studied a modified version of the BM slicing condition
that has to important features: 1) it guarantees that if a spacetime
is static or stationary, and one starts the evolution in a coordinate
system in which the metric coefficients are already time independent,
then they will remain time independent during the subsequent
evolution, and 2) the modified condition is naturally adapted to the
use of a densitized lapse as a fundamental variable.

By analyzing this modified BM condition written in covariant form on
an arbitrary background spacetime, we have also shown that it is
strongly hyperbolic for $f(\alpha)>0$, just as the original BM
condition was.  Moreover, we have found that the characteristic speeds
of the original BM condition are not modified.  Finally, we have shown
that as long as the shift vector remains regular, the modified BM
condition avoids both focusing singularities and gauge shocks under
the same conditions as the original BM condition.
 
Because of these results we believe that the modified BM condition
might be just as useful as the original BM condition for evolving
strongly gravitating systems, while at the same time having the extra
benefits described above.  We plan to carry out numerical experiments
to test this, but since the modified BM condition leads to a different
slicing of spacetime for different choices of shift, such experiments
will require first that one studies different shift conditions.  We
are currently working on this issue, and will report our findings in a
future work.


\acknowledgments

We thank Olivier Sarbach for useful discussions.  This work was
supported in part by CONACyT through the repatriation program and
grants 149945, 32551-E and J32754-E, by DGAPA-UNAM through grants IN112401 and
IN122002, and by DGEP-UNAM through a complementary grant.


\appendix*
\section{}
\label{sec:appendix}

While studying the hyperbolicity of the modified BM slicing condition in
Section~\ref{sec:hyperbolicity} we found, surprisingly, that two of
the eigenvalues associated with this modified condition are identical
with the eigenvalues associated with the standard BM condition found
in Ref.~\cite{Alcubierre02b}.  The analysis of
Section~\ref{sec:hyperbolicity} was done in a covariant way, which
required the introduction of an evolution equation for the shift that
stated the fact that the shift has to be parallel to the
hypersurfaces. The introduction of the shift equation made the
analysis considerably more complicated and makes it difficult to see
why the eigenvalues should remain equal.  Here we will do a simple
analysis in the 3+1 coordinate frame for the particular case of one
spatial dimension to try to understand why the characteristic speeds
are not modified.

Let us then consider first the standard BM slicing condition in one
spatial dimension.  The evolution equation for the lapse has the form
\begin{equation}
\partial_t \alpha = \beta \partial_x \alpha - \alpha^2 f K \; .
\end{equation}
If we now introduce the first order quantity $A := \partial_x \alpha$,
we can rewrite this equation as the system
\begin{eqnarray}
\partial_t \alpha &=& \beta A - \alpha^2 f K \; , \\
\partial_t A &=& \partial_x \left( \beta A - \alpha^2 f K \right) \; ,
\end{eqnarray}
with $K$ the trace of the extrinsic curvature.   On the other hand,
from the ADM equations we find the following evolution equation for $K$
\begin{equation}
\partial_t K \simeq \partial_x \left( \beta K - A / \gamma \right) \; , 
\end{equation}
where the symbol $\simeq$ denotes equal up to principal part, and
where $\gamma$ is the one-dimensional coordinate volume element.
Notice that we are considering the shift $\beta$ to be a prescribed
function of space and time.

The Jacobian matrix associated with the evolution equations for
$(A,K)$ is then
\begin{equation}
M = \left(
\begin{array}{cc}
- \beta & \alpha^2 f \\
1/\gamma & - \beta 
\end{array}
\right) \; .
\label{eq:jacobian1}
\end{equation}
Notice that the overall sign chosen here is the one obtained by moving
the spatial derivatives to the right hand side of the equations, as
this is the sign we need if we want to associate the eigenvalues with
characteristic speeds. The associated eigenvalues are
\begin{equation}
\lambda_\pm = - \beta \pm \alpha \left( f / \gamma \right)^{1/2} \; ,
\end{equation}
with corresponding eigenvectors
\begin{equation}
v_\pm = \left( \pm \alpha \left( f \gamma \right)^{1/2}, 1 \right) \; . 
\end{equation}
We then see that evolution equations for the pair $(A,K)$ form a
strongly hyperbolic system as long as $f>0$.

Let us now consider the modified BM slicing condition.  The evolution
equation for the lapse now becomes
\begin{equation}
\partial_t \alpha = - \alpha f \left[ \alpha K - \partial_x \beta
- \left( \beta / 2 \gamma \right) \; \partial_x \gamma \right] \; .
\end{equation}
Notice how this now includes a spatial derivative of the coordinate
volume element, which can not be considered as a source since by
construction the time derivative of the lapse is proportional to the
time derivative of the volume element.  If we now define $G:=
\partial_x \gamma$, we can rewrite the evolution equations for
$\alpha$ and $\gamma$ as the system
\begin{eqnarray}
\partial_t \alpha &=& -\alpha^2 f K + \frac{\alpha f \beta}{2 \gamma} \; G
+ \alpha f \partial_x \beta \; , \\
\partial_t \gamma &=& - 2 \alpha \gamma K + \beta G
+ 2 \gamma \partial_x \beta \; , \\
\partial_t A &\simeq& \partial_x \left( -\alpha^2 f K
+ \frac{\alpha f \beta}{2 \gamma} \; G \right) \; , \\
\partial_t G &\simeq& \partial_x \left( - 2 \alpha \gamma K + \beta G
\right) \; .
\end{eqnarray}
We now see that the Jacobian matrix associated with the system
$(A,G,K)$ takes the form
\begin{equation}
M = \left(
\begin{array}{ccc}
0 & - \alpha f \beta / 2 \gamma & \alpha^2 f \\
0 & - \beta & 2 \alpha \gamma \\
1/\gamma & 0 & - \beta 
\end{array}
\right) \; .
\label{eq:jacobian2}
\end{equation}
The eigenvalues of this matrix are easily found to be
\begin{eqnarray}
\lambda_0 &=& 0 \; , \\
\lambda_\pm &=& - \beta \pm \alpha \left( f / \gamma \right)^{1/2} \; ,
\end{eqnarray}
with associated eigenvectors
\begin{eqnarray}
v_0 &=& \left( \beta^2 \gamma , 2 \alpha \gamma , \beta \right) \; , \\
v_\pm &=& \left( \pm \alpha \left( f \gamma \right)^{1/2} ,
\pm 2 \; \gamma^{3/2} / f^{1/2} , 1 \right) \; .
\end{eqnarray}
We can again see that two of the eigenvalues corresponding to the
modified BM slicing condition coincide with the eigenvalues of the
standard BM condition.  However, the reason for this is now much
easier to see.  Notice that the Jacobian matrix~(\ref{eq:jacobian2})
has two rows that are multiples of each other (row two can be obtained
from row one by multiplying with $2 \gamma / \alpha f$).  This means
that we can define a new variable $\Sigma := G - (2 \gamma / \alpha f)
A$ that will evolve only through lower order terms.  We can now make a
change of variables from $(A,G,K)$ to $(A,\Sigma,K)$, by replacing $G$
with $\,\Sigma + (2 \gamma / \alpha f) A$.  Since $\Sigma$ evolves only
through lower order terms, its derivatives can be treated as source
terms in the hyperbolicity analysis.  It is then easy to see that the
Jacobian matrix for the reduced system $(A,K)$ is identical to the
Jacobian matrix associated with the standard BM slicing condition
given in Eq.~(\ref{eq:jacobian1}).

The important observation is that even though the modified BM
slicing condition replaces the Lie derivative of $\alpha$ with respect
to the shift with the divergence of the shift, since this divergence
includes the Lie derivative of $\gamma$, and the time derivative of
$\alpha$ and $\gamma$ are multiples of each other, we recover in the
end precisely the same characteristic speeds $- \beta \pm \alpha
\left( f / \gamma \right)^{1/2}$.


\bibliographystyle{bibtex/apsrev}
\bibliography{bibtex/referencias}


\end{document}